\documentclass[letterpaper]{article}
\usepackage{aaai2026}
\usepackage{times}
\usepackage{enumitem}
\usepackage{multirow}
\usepackage{linguex}
\usepackage{braket}
\usepackage{enumitem}
\usepackage{xcolor}
\usepackage{booktabs}
\usepackage{amsmath,amssymb,amsfonts}
\usepackage{todonotes}
\usepackage{adjustbox}
\usepackage{multirow}
\usepackage{makecell}
\usepackage{graphicx}
\usepackage{ragged2e}
\usepackage{multirow}
\usepackage{braket}
\usepackage{amsmath}
\usepackage{amssymb}
\usepackage{amsthm}

\newtheorem{definition}{Definition}
\usepackage{helvet}
\usepackage{courier}
\usepackage{natbib}
\usepackage{caption}
\usepackage{algorithm}
\usepackage{algorithmic}

\usepackage{newfloat}
\usepackage{listings}
\DeclareCaptionStyle{ruled}{labelfont=normalfont,labelsep=colon,strut=off} 
\floatstyle{ruled}
\newfloat{listing}{tb}{lst}{}
\floatname{listing}{Listing}
\newcommand{\modelname}{\texttt{QuanTaxo}}
\newcommand{\mixture}{\texttt{Quant-Mix}}
\newcommand{\superposition}
{\texttt{Quant-Sup}}
\title{\modelname: A Quantum Approach to Self-Supervised Taxonomy Expansion}
\author{
    Sahil Mishra\textsuperscript{\rm 1},
    Avi Patni\textsuperscript{\rm 2},
    Niladri Chatterjee\textsuperscript{\rm 2},
    Tanmoy Chakraborty\textsuperscript{\rm 1}
}
\affiliations{
    \textsuperscript{\rm 1}Department of Electrical Engineering, \textsuperscript{\rm 2}Department of Mathematics\\Indian Institute of Technology Delhi\\
    sahil.mishra@ee.iitd.ac.in, patni.avi@gmail.com, \{niladri, tanchak\}@iitd.ac.in
}

\usepackage{bibentry}

\begin{document}

\maketitle

\begin{abstract}
A taxonomy is a hierarchical graph containing knowledge to provide valuable insights for various web applications. However, the manual construction of taxonomies requires significant human effort. As web content continues to expand at an unprecedented pace, existing taxonomies risk becoming outdated, struggling to incorporate new and emerging information effectively. As a consequence, there is a growing need for dynamic taxonomy expansion to keep them relevant and up-to-date. Existing taxonomy expansion methods often rely on classical word embeddings to represent entities. However, these embeddings fall short of capturing hierarchical polysemy, where an entity's meaning can vary based on its position in the hierarchy and its surrounding context. To address this challenge, we introduce \modelname, a quantum-inspired framework for taxonomy expansion that encodes entities in a Hilbert space and models interference effects between them, yielding richer, context-sensitive representations. Comprehensive experiments on five real-world benchmark datasets show that \modelname\ significantly outperforms classical embedding models, achieving substantial improvements of 12.3\% in accuracy, 11.2\% in Mean Reciprocal Rank (MRR), and 6.9\% in Wu \& Palmer (Wu\&P) metrics across nine classical embedding-based baselines.
\end{abstract}
\section{Introduction}
\label{sec:Intro}

Taxonomy is a hierarchically structured knowledge graph designed to portray the hypernymy (``is-a'') relationship between concepts, showing how broad categories subsume more specific ones. Therefore, it serves as core infrastructure for many online applications by efficiently indexing and organizing knowledge. For example, Amazon \cite{mao2020octet} and Alibaba \cite{karamanolakis_txtract_2020} use taxonomies to power e-commerce search and browsing, while Pinterest leverages them for content recommendation and advertisement targeting \cite{owl,manzoor2020expanding}.


\begin{figure}[!t]
\centering
\includegraphics[width=0.9\columnwidth]{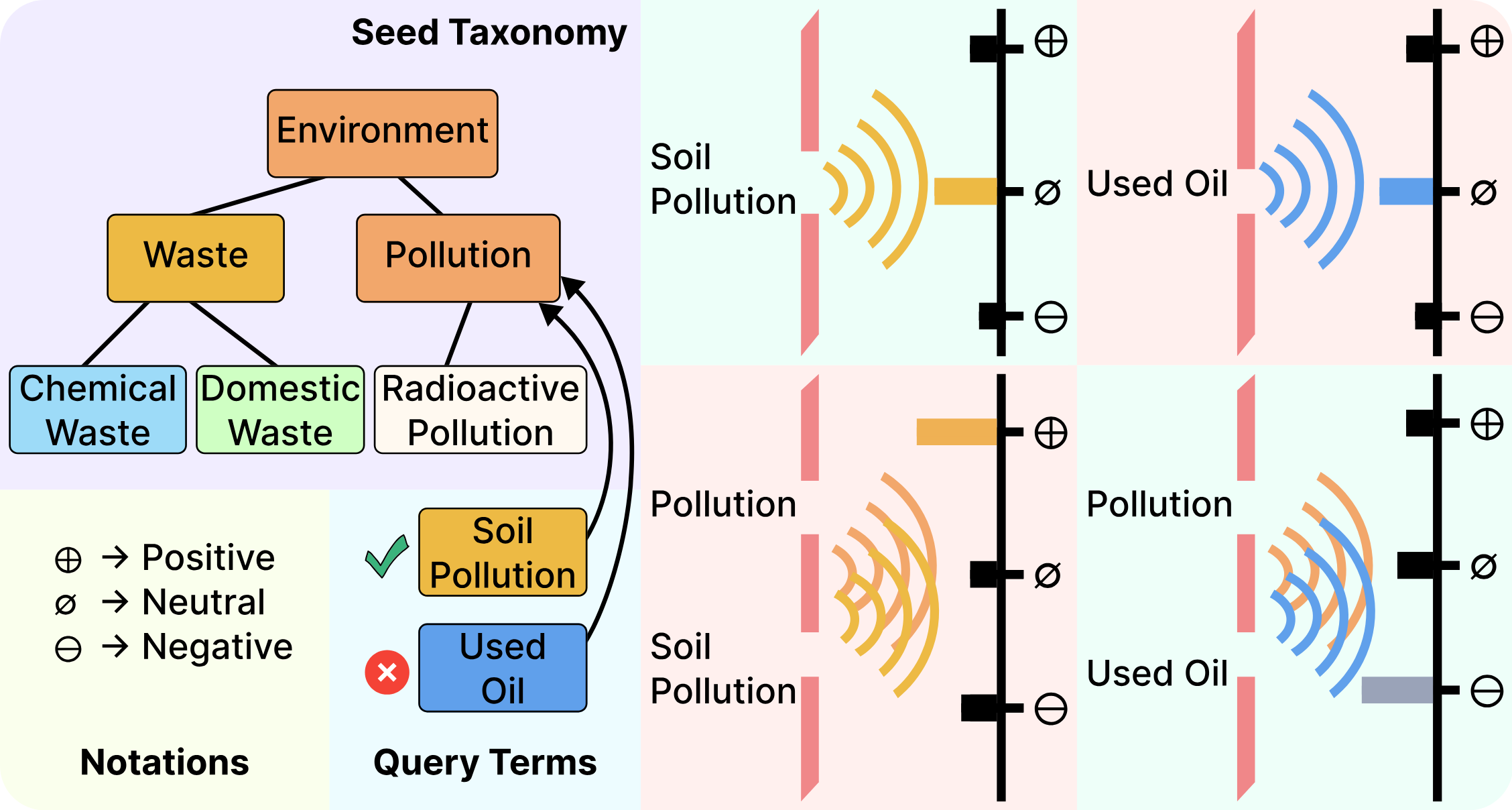}
\caption{An illustration of taxonomy expansion using \modelname\ on the \textit{Environment} seed taxonomy, where the query terms \textit{Soil Pollution} and \textit{Used Oil} are to be inserted.}
\label{fig:intro}
\end{figure}

Traditional taxonomies are typically constructed from scratch by domain experts, making the process slow and labor-intensive. Early automation efforts used unsupervised methods such as graph pruning \cite{velardi-etal-2013-ontolearn}, hierarchical clustering \cite{zhang2018taxogen}, and topic modeling \cite{song2016automatic, wang2013phrase} to induce taxonomies from raw data, but they often fail to match the structure and coherence of expert-designed taxonomies. As data continuously evolves, there is a growing need to incorporate new concepts into existing taxonomies. To address this, we focus on \textit{taxonomy expansion} —- the task of integrating new concepts, or \textit{query nodes}, into an existing \textit{seed taxonomy} by placing them under suitable \textit{anchor nodes}. For example, as shown in Fig.~\ref{fig:intro}, \textit{Soil Pollution} can be accurately placed under \textit{Pollution} to maintain structural consistency.

Early work on taxonomy expansion used self-supervision, relying on lexical patterns or distributional embeddings to learn parent-child relationships from a small seed taxonomy \cite{jurgens-pilehvar-2015-reserating,snow2004learning}. However, these methods struggled with limited data and underused structural information. More recent approaches improve this by using paths \cite{liu2021temp,jiang2022taxoenrich} or local graphs \cite{wang2021enquire,berant-etal-2015-efficient} to better model hierarchy. Others leverage hyperbolic geometry \cite{ganea2018hyperbolic,nickel2017poincare} or box embeddings \cite{abboud2020boxe,jiang2023single} for more expressive hierarchical representations.

Most existing methods represent taxonomy entities using classical word embeddings and infer parent–child relations by measuring similarity between their vectors. However, such embeddings struggle to capture nuanced, compositional meaning. For instance, ``fish'' and ``drown'' may appear similar because they co-occur in sentences like ``\textit{Fish can swim in water where others would drown}.'' Yet, these embeddings fail to capture the negative connotation - ``\textit{Fish cannot drown}.'' These limitations call for more expressive representations that model hierarchical and contextual subtleties. This motivates quantum-inspired embeddings, which use principles such as superposition and entanglement to better encode such complex semantics.

To show the importance of quantum embeddings, we model the parent-child hierarchy in the taxonomy by treating individual entities as having limited standalone significance, while their superposition reveals the degree of relatedness between them. Specifically, as illustrated in Fig. \ref{fig:intro}, we draw inspiration from Fraunhofer's double-slit experiment in Quantum Physics \cite{born2013principles}. In this analogy, the two slits represent the parent-child relationship between taxonomy entities. When only one of the slits is opened, the wave corresponding to the individual word passes through, registering on the detection screen as a neutral entity, as seen in cases like ``Soil Pollution'' and ``Used Oil.'' However, when both slits are opened, the superposition of the waves from both words reveals the nature of their relationship, which is positive between ``Pollution'' and ``Soil Pollution'', but negative between ``Pollution'' and ``Used Oil.''

We introduce \modelname \footnote{Code: https://github.com/sahilmishra0012/QuanTaxo/}, the \textbf{Quan}tum \textbf{Taxo}nomy Expansion Framework that leverages the superposition principle from quantum physics to represent hierarchical polysemy in a taxonomy. Our key contributions are as follows:

We first construct training data using a self-supervised framework based on the seed taxonomy. Each \(\langle\)parent, child\(\rangle\) pair from the taxonomy serves as a positive example, while negative samples are created by pairing the child with non-ancestral nodes. As illustrated in Fig.~\ref{fig:intro}, \(\langle\)Environment, Waste\(\rangle\) and \(\langle\)Pollution, Radioactive Pollution\(\rangle\) are positive samples, whereas \(\langle\)Pollution, Chemical Waste\(\rangle\) and \(\langle\)Chemical Waste, Radioactive Pollution\(\rangle\) are negatives.

Secondly, we propose a quantum modeling framework to represent \(\langle\)parent, child\(\rangle\) pairs in a complex probabilistic Hilbert space. Inspired by Fraunhofer's double-slit experiment, we adopt an entanglement-based approach to superimpose parent and child embeddings, yielding quantum entity representations via density matrices. The framework is based on two core hypotheses: (i) a word is a linear combination of latent concepts with complex weights, and (ii) multiple words form a complex superposition of their states. Accordingly, we introduce two variants: (i) \superposition, which models entities as linear combinations of latent concepts, and (ii) \mixture, which models them as weighted mixtures of word states.

Thirdly, we propose a joint representation framework to quantify the relatedness between parent and child entities in the taxonomy. This involves superimposing their quantum representations to form a composite representation that captures their degree of ``entanglement'' or interconnectedness. From this, we extract specialized ``entangled features'' reflecting relational coherence. We assess relatedness using mathematical properties such as the trace and diagonal elements of the joint matrix—the trace captures cumulative shared attributes, while the diagonal highlights individual contributions and structural alignment within the hierarchy. Ablation studies comparing complex and real embeddings reveal why the complex space is essential, and comprehensive experiments on five benchmarks against nine strong baselines show that QuanTaxo lifts performance by an average of 12.3\% in accuracy, 11.2\% in MRR, and 6.9\% in Wu\&P metrics, confirming its effectiveness for taxonomy expansion.

\section{PRELIMINARIES}
\label{sec:prelim}
\subsection{Hilbert Space}
In quantum probability theory \cite{nielsen2010quantum, von2018mathematical}, quantum systems are modeled within a Hilbert space \(\mathbb{H}^n\), a complex vector space where states are represented as vectors (or density operators), and probabilities are computed from their inner products.

We follow the Dirac notation commonly used in quantum theory, where a state vector \(\vec{\psi} \in \mathbb{C}^n\) is denoted as a ket \(\ket{\psi}\), and its transpose as a bra \(\bra{\psi}\). The inner and outer products of unit vectors \(\vec{u}\) and \(\vec{v}\) are written as \(\braket{u|v}\) and \(\ket{u}\bra{v}\), respectively. A vector \(\ket{\psi}\) can be expressed as a superposition of basis vectors,
\begin{equation}
    \ket{\psi} = \sum_{i=1}^{n}{a_je^{i\phi_j}\ket{e_j}},
    \label{eq:superposition}
\end{equation}
where \(a_j e^{i\phi_j}\) is the complex-valued probability amplitude for the \(j^{\text{th}}\) basis vector \(\ket{e_j}\). Here, \(a_j \geq 0\) are real amplitudes satisfying the normalization \(\sum_j a_j^2 = 1\), and \(\phi_j \in [-\pi, \pi]\) are the phase angles. Each amplitude can also be expressed in Euler form as \(a_j \cos \phi_j + i a_j \sin \phi_j\), and is computed via the inner product: \(a_j e^{i\phi_j} = \braket{e_j|\psi}\).

Further, the projection measurement is computed as \(p(e_j \mid \psi) = a_j^2 = \bigl|\langle e_j \mid \psi \rangle\bigr|^2\), where \(p\left(e_j \mid \psi\right)\) represents the probability of the quantum event \(\ket{e_1}\), given the quantum state \(\ket{\psi}\). The vector \(\ket{\psi}\) in Eq. \ref{eq:superposition} represents a word as a combination of sememes \cite{goddard1994semantic} \footnote{Sememes are also referred to as latent concepts. Each word is a combination of \textit{n} latent concepts.}, which are the fundamental, indivisible semantic components of word meanings in a language. For instance, the word ``robot'' can be composed of sememes like ``machine'', ``automation'', ``technology'' and ``artificial.'' The complex phases $\{\phi_j\}_{j=0}^n$ capture \textit{quantum interference} between words. For example, given two words, $w_k$ and $w_p$ with complex amplitudes $a_j^{(k)} e^{i\phi_j^{(k)}}$ and $a_j^{(p)} e^{i\phi_j^{(p)}}$ respectively for the sememe $e_j$, their combination affects the probability of being in state $e_j$ as follows \(\left|a_j^{(k)} e^{i\phi_j^{(k)}}+a_j^{(p)} e^{i\phi_j^{(p)}}\right|^2 = \left|a_j^{(k)}\right|^2 + \left|a_j^{(p)}\right|^2 +  2a_j^{(k)}a_j^{(p)}\cos\left(\phi_j^{(k)}-\phi_j^{(p)}\right)\), where the term \(2a_j^{(k)}a_j^{(p)}\cos\left(\phi_j^{(k)}-\phi_j^{(p)}\right)\) represents the interference between \(w_k\) and \(w_p\).

\subsection{Sentence Representation}
\label{subsec:sentencerepresentation}
A sentence is formed as a combination of words, with each existing in a superposition of underlying sememes.  Therefore, the sentence \( S \) is a non-classical combination of these sememes. Mathematically, it is represented by an \( n \times n \) density matrix \( \rho \), which is positive semi-definite (\( \rho \geq 0 \)) and with a unit trace (\( \text{Tr}(\rho) = 1 \)). The diagonal elements of \( \rho \) indicate the contribution of individual concepts, while off-diagonal elements capture their quantum-like correlations between them.

In quantum probabilistic space, a sentence can be represented in two ways: \textit{superposition} and \textit{mixture}. In the \textit{superposition} representation, the sentence exists simultaneously in multiple potential states as a combination of latent concepts, capturing the inherent uncertainty and overlap of interpretations, akin to a quantum particle that exists in multiple states. The corresponding density matrix for the sentence \( S \) is,
\begin{equation}
\rho = \ket{S}\bra{S},
\label{eq:sup}
\end{equation}
where sentence \( S \) is represented (using Eq. \ref{eq:superposition}) as \(\ket{S} = \sum_{i=1}^{n}{a_je^{i\phi_j}\ket{e_j}}\).

While in the \textit{mixture} representation, a sentence is a combination of different word states, where each interpretation is considered as a distinct possibility, weighted by its corresponding probability. Unlike superposition, where states coexist as latent concepts, the mixture approach assigns a classical probability distribution over possible word states. The density matrix is computed as,
\begin{equation}
    \rho = \sum_{i=1}^{n}{\lambda_i\ket{\psi_i}\bra{\psi_i}},
    \label{eq:mix}
\end{equation}
where \(\lambda_i\) represents the weight of the word \(\psi_i\) with \(\sum_{i=1}^{n}\lambda_i=1\).

\begin{figure}[!t]
    \centering
\includegraphics[width=0.45\textwidth]{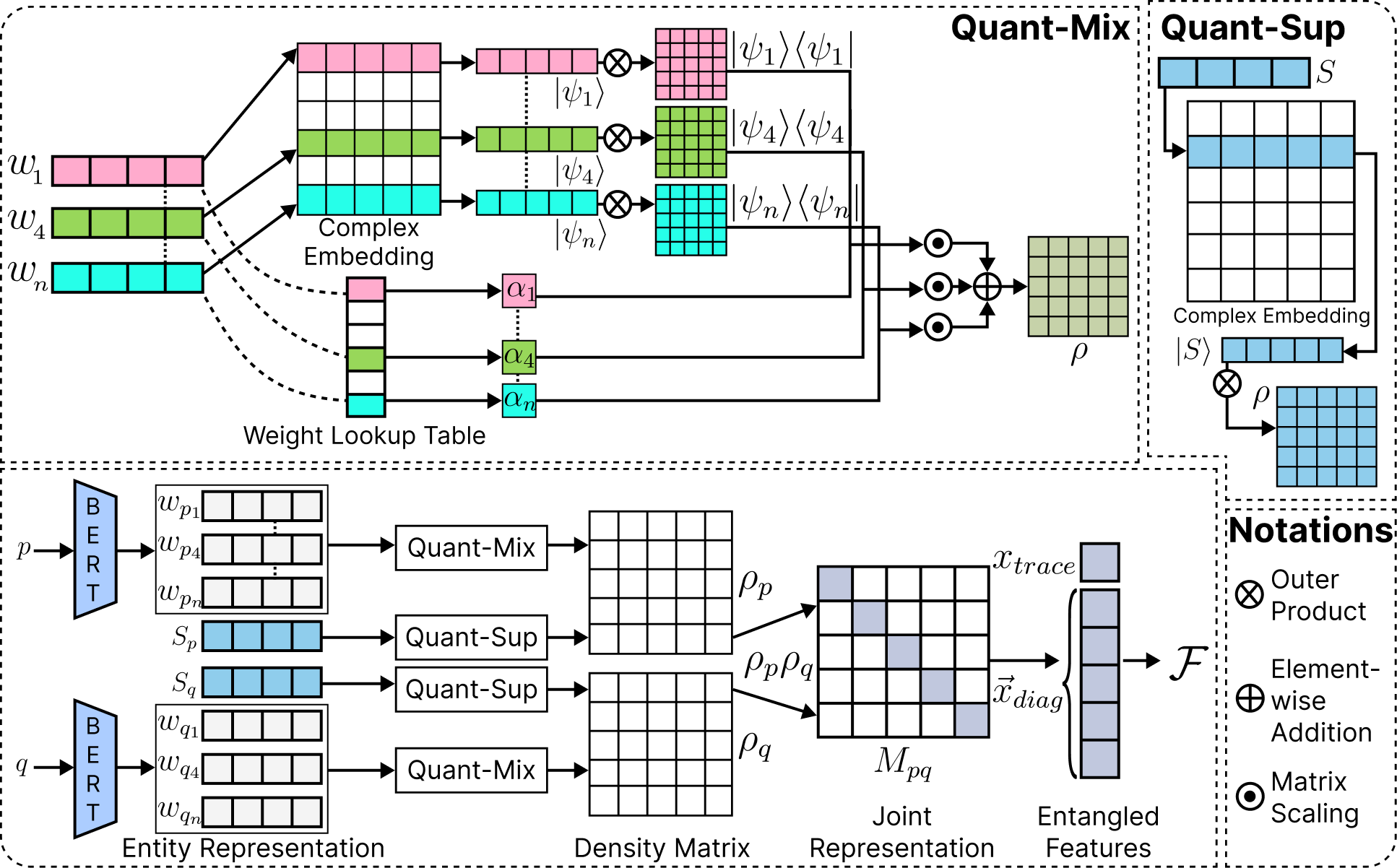}
    \caption{An illustration of the \modelname\ framework. Given parent and query entities, their complex representations are generated, followed by outer products on the word and sentence embeddings to generate quantum representations. The joint representation of parent and child entities is then computed to score the semantic relationship.}
    \label{fig:model}
\end{figure}

\subsection{Taxonomy Expansion}
\begin{definition}
    \textbf{Taxonomy:} A taxonomy \(\mathcal{T}^o=(\mathcal{N}^o,\mathcal{E}^o)\) is a tree-like directed acyclic graph, where each node \(n \in \mathcal{N}^o\) represents a concept and each edge \(\langle n_p, n_c \rangle \in \mathcal{E}^o\) denotes a "parent-child" relationship between nodes \(n_p\) and \(n_c\).
\end{definition}
\begin{definition}
    \textbf{Taxonomy Expansion:} Given a seed taxonomy \(\mathcal{T}^o=(\mathcal{N}^o,\mathcal{E}^o)\) and a set of emerging concepts \(C\), the task is to update the seed taxonomy to \(\mathcal{T}=(\mathcal{N}^o \cup \mathcal{C},\mathcal{E}^o \cup \mathcal{R})\), where \(\mathcal{R}\) is the set of newly created relationships linking existing entities \(\mathcal{E}^o\) with emerging entities \(C\). Since surface names of entities alone lack true semantics, entity descriptions \(D\) are used to augment representations. Moreover, during inference, query node \(q \in \mathcal{C}\) identifies its best-suited parent node \(n_p \in \mathcal{N}^o\) by maximizing the matching score ($n_p=\arg \max _{a \in \mathcal{N}^0} f(a, q)$).
\end{definition}

\section{The \modelname\ Framework}
\label{sec:model}
This section presents \modelname, an end-to-end neural network implemented in the Hilbert space (c.f Figure \ref{fig:model}).

\subsection{Complex-valued Entity Representation}
\label{subsec:complex}
We start with computing the complex-valued representation of the entities. Given a candidate term \(n_p \in \mathcal{N}^o\) and a query term \(n_q\) (\(n_q \in \mathcal{N}^o\) during training or \(n_q \in \mathcal{C}\) while inference), we compute their real-valued vector representations from their surface names and descriptions and then project them into complex space.

\noindent \textbf{Real-valued Entity Projection:} Taxonomy nodes, defined by surface names and descriptions \(D\), are encoded into real-valued vectors using pretrained models like BERT \cite{devlin2018bert}, in preparation for projection into complex space \(\mathbb{C}^n\). Each entity \(n_e \in \{n_p, n_q\}\) is formatted as \(d_e = \texttt{[CLS]}\texttt{sur}(n_e)\texttt{[SEP]}{D}(n_e)\texttt{[SEP]}\), where \(\texttt{sur}(n_e)\) and \(\texttt{D}(n_e)\) denote the entity's surface name and description, respectively, and \(\texttt{[CLS]}, \texttt{[SEP]}\) are BERT-specific tokens.

The textual input \(d_e\) is encoded by BERT as \(z_e = \text{BERT}(d_e)\), where \(z_e\) contains the final-layer embeddings of all tokens. We extract the sentence embedding \(S = z_{e_{[0]}}\) from the \(\texttt{[CLS]}\) token for the \superposition\, and token embeddings \(\{\psi\}_{i=0}^m = z_{e_{[1:m]}}\) for the \mixture.

\noindent \textbf{Complex-valued Entity Projection:} We project real-valued embeddings into the Hilbert space \(\mathbb{H}^n = \mathbb{C}^n\) using a complex phase projector parameterized by amplitude \(A\), phase \(\Phi\), and a learnable weight table \(\Lambda\). Given a real-valued vector \(x\), the projection is defined as \(\texttt{ComplexProjector}(x) = A(x) \odot e^{i \Phi(x)}\), where \(A(x) = \|x\|_2\) computes the amplitude, and \(\Phi(x) = f_\phi(x)\) applies a linear transformation to generate the phase. To maintain a probabilistic interpretation, amplitudes are normalized such that \(\sum_{j=0}^{n} a_j = 1\), \(\forall a_j \in A\). The complex word embedding is then \(\ket{\psi_i} = \texttt{ComplexProjector}(\psi_i)\), combining amplitude \(a\) and phase \(\phi\) as per Eq.~\ref{eq:superposition}. The weights \(\Lambda = \{\alpha_i\}_{i=0}^n\) modulate semantic importance in the \mixture\ network (Section~\ref{subsec:quant}). For the \superposition\ network, the complex embedding is \(\ket{S} = \texttt{ComplexProjector}(S)\).

\begin{table}[!t]
\centering
\resizebox{0.8\linewidth}{!}{
\begin{tabular}{lrrr}
\toprule
\textbf{Dataset} & \boldmath${|\mathcal{N}^0|}$ & \boldmath${|\mathcal{E}^0|}$ & \boldmath${|D|}$ \\
\midrule
SemEval-Env      & 261 & 261 & 6 \\
SemEval-Sci      & 429 & 452 & 8 \\
MAG-CS-WIKI      & 25,170 & 40,314 & 6 \\
MAG-PSY-WIKI     & 10,671 & 14,080 & 6 \\
WordNet          & 20.5 & 19.5 & 3 \\
\bottomrule
\end{tabular}
}
\caption{Statistics of benchmark datasets. ${|\mathcal{N}^0|}$ and ${|\mathcal{E}^0|}$ denote the number of nodes and edges in the seed taxonomy, while ${|D|}$ is the taxonomy depth. For WordNet, values are averaged across 114 sub-taxonomies.}
\label{table:dataset}
\end{table}

\begin{table*}[t]
\centering
\renewcommand{\arraystretch}{1.3}
\setlength{\tabcolsep}{2.5pt}
\scalebox{0.89}{
\begin{tabular}{|l|ccc|ccc|ccc|ccc|ccc|}
\hline
\textbf{Dataset} & \multicolumn{3}{c|}{\textbf{SemEval16-Env}} & \multicolumn{3}{c|}{\textbf{SemEval16-Sci}} & \multicolumn{3}{c|}{\textbf{MAG-WIKI-CS}} & \multicolumn{3}{c|}{\textbf{WordNet}} & \multicolumn{3}{c|}{\textbf{MAG-WIKI-PSY}} \\ \hline
\textbf{Metric} & \textbf{Acc} & \textbf{MRR} & \textbf{Wu\&P} & \textbf{Acc} & \textbf{MRR} & \textbf{Wu\&P} & \textbf{Acc} & \textbf{MRR} & \textbf{Wu\&P} & \textbf{Acc} & \textbf{MRR} & \textbf{Wu\&P} & \textbf{Acc} & \textbf{MRR} & \textbf{Wu\&P} \\ \hline
BERT+MLP & 12.6$^\text{1.1}$ & 23.9$^\text{1.6}$ & 48.3$^\text{0.8}$ & 12.2$^\text{1.7}$ & 19.7$^\text{1.4}$ & 45.1$^\text{1.1}$ & 8.3$^\text{2.4}$ & 15.7$^\text{1.5}$ & 36.2$^\text{1.3}$ & 9.2$^\text{1.2}$ & 17.4$^\text{1.3}$ & 43.5$^\text{0.4}$ & 7.8$^\text{1.4}$ & 15.5$^\text{1.1}$ & 37.1$^\text{0.6}$ \\
TAXI & 18.5$^\text{1.3}$ & N/A & 47.7$^\text{0.4}$ & 13.8$^\text{1.4}$ & N/A & 33.1$^\text{0.7}$ & 17.3$^\text{1.6}$ & N/A & 42.3$^\text{0.5}$ & 11.5$^\text{1.8}$ & N/A & 38.7$^\text{0.7}$ & 12.9$^\text{1.4}$ & N/A & 39.7$^\text{0.6}$ \\
Arborist & 9.9$^\text{3.9}$ & 27.2$^\text{3.5}$ & 43.4$^\text{1.6}$ & 19.1$^\text{5.0}$ & 32.7$^\text{3.2}$ & 47.8$^\text{1.8}$ & 15.6$^\text{3.7}$ & 34.4$^\text{2.9}$ & 46.3$^\text{1.8}$ & 12.5$^\text{3.3}$ & 27.7$^\text{2.1}$ & 52.0$^\text{1.2}$ & 15.1$^\text{2.8}$ & 28.1$^\text{1.9}$ & 44.9$^\text{1.5}$ \\
TaxoExpan & 10.7$^\text{4.1}$ & 28.7$^\text{3.8}$ & 48.5$^\text{1.7}$ & 24.2$^\text{5.4}$ & 40.3$^\text{3.3}$ & 55.6$^\text{1.9}$ & 16.8$^\text{4.1}$ & 34.9$^\text{3.2}$ & 47.7$^\text{1.3}$ & 17.3$^\text{3.5}$ & 31.1$^\text{2.3}$ & 57.6$^\text{1.8}$ & 15.9$^\text{2.2}$ & 30.4$^\text{1.8}$ & 51.2$^\text{1.1}$ \\
TMN & 34.6$^\text{3.0}$ & 41.7$^\text{4.4}$ & 53.6$^\text{3.5}$ & 32.6$^\text{2.4}$ & 46.1$^\text{1.9}$ & 65.3$^\text{1.5}$ & 24.4$^\text{3.2}$ & 39.7$^\text{2.3}$ & 53.1$^\text{1.6}$ & 20.3$^\text{1.9}$ & 35.9$^\text{1.5}$ & 54.7$^\text{1.3}$ & 19.7$^\text{2.0}$ & 34.8$^\text{1.7}$ & 51.9$^\text{0.3}$ \\
STEAM & 34.1$^\text{3.4}$ & 44.3$^\text{2.1}$ & 65.2$^\text{1.4}$ & 34.8$^\text{4.5}$ & 50.7$^\text{2.5}$ & 72.1$^\text{1.7}$ & 25.9$^\text{4.0}$ & 41.3$^\text{2.5}$ & 53.7$^\text{0.4}$ & 21.4$^\text{2.8}$ & 38.2$^\text{1.4}$ & 59.8$^\text{1.3}$ & 23.5$^\text{3.1}$ & 36.3$^\text{1.6}$ & 53.3$^\text{1.2}$ \\
BoxTaxo & 32.3$^\text{5.8}$ & 45.7$^\text{3.2}$ & 73.1$^\text{1.2}$ & 26.3$^\text{4.5}$ & 41.1$^\text{3.1}$ & 61.6$^\text{1.4}$ & 23.5$^\text{5.4}$ & 35.7$^\text{4.1}$ & 49.2$^\text{1.7}$ & 22.3$^\text{3.1}$ & 35.7$^\text{2.7}$ & 58.6$^\text{1.2}$ & 21.1$^\text{3.4}$ & 34.9$^\text{2.9}$ & 54.0$^\text{1.3}$ \\ 
Musubu & 42.3$^\text{3.2}$ & 57.1$^\text{1.4}$ & 64.4$^\text{0.7}$ & 44.5$^\text{2.3}$ & 59.7$^\text{1.6}$ & 75.2$^\text{1.2}$ & 28.5$^\text{2.9}$ & 37.1$^\text{2.0}$ & 51.9$^\text{0.7}$ & \underline{25.3}$^\text{4.9}$ & 36.1$^\text{2.9}$ & 61.2$^\text{0.9}$ & 25.7$^\text{4.4}$ & 35.0$^\text{2.3}$ & 57.1$^\text{1.0}$ \\
TEMP & 45.5$^\text{8.6}$ & \underline{59.1}$^\text{6.3}$ & \underline{77.3}$^\text{2.8}$ & 43.5$^\text{7.8}$ & 57.5$^\text{5.6}$ & 76.3$^\text{1.5}$ & 31.8$^\text{1.4}$ & 45.3$^\text{2.1}$ & 57.8$^\text{1.2}$ & 24.6$^\text{5.1}$ & 37.5$^\text{4.6}$ & 61.2$^\text{2.3}$ & \underline{25.9}$^\text{2.1}$ & \underline{38.9}$^\text{1.9}$ & \underline{59.7}$^\text{0.3}$ \\
\hline
\superposition\ & \textbf{49.2$^\text{2.1}$} & \textbf{59.5$^\text{1.2}$} & \textbf{79.1$^\text{0.3}$} & \textbf{57.3$^\text{2.8}$} & \textbf{67.4$^\text{1.2}$} & \textbf{82.1$^\text{0.4}$} & \textbf{35.7$^\text{2.2}$} & \textbf{51.8$^\text{1.6}$} & \textbf{61.4$^\text{0.6}$} & \textbf{25.8$^\text{1.1}$} & \textbf{41.8$^\text{0.6}$} & \textbf{71.0$^\text{0.3}$} & \textbf{27.3$^\text{1.4}$} & \textbf{43.2$^\text{0.7}$} & \textbf{61.0$^\text{0.4}$} \\
\mixture\ & \underline{46.6}$^\text{2.1}$ & 57.9$^\text{1.2}$ & 76.1$^\text{0.1}$ & \underline{52.4}$^\text{1.7}$ & \underline{64.1}$^\text{1.1}$ & \underline{77.6}$^\text{0.6}$ & \underline{33.2}$^\text{1.9}$ & \underline{47.7}$^\text{1.1}$ & \underline{59.2}$^\text{0.7}$ & 22.1$^\text{1.3}$ & \underline{39.6}$^\text{0.8}$ & \underline{69.8}$^\text{0.4}$ & 23.5$^\text{1.5}$ & 38.4$^\text{1.0}$ & 59.4$^\text{0.5}$ \\ \hline
\end{tabular}}
\caption{Performance comparison between \modelname\ and baseline methods. Results for each method are presented as \( \textbf{mean}^\textbf{std\text{-}dev} \) in percentage across three runs with three random seeds. The best performance is marked in bold, while the best baseline is underlined. As TAXI \cite{panchenko2016taxi} outputs the taxonomy as a whole, it cannot produce MRR values.}
\label{table:results}
\end{table*}

\subsection{Quantum Representation}
\label{subsec:quant}
Quantum embeddings of entities are modeled as density matrices (Section~\ref{subsec:sentencerepresentation}). The \superposition\ and \mixture\ networks compute superposition and mixture-based representations, respectively. In the mixture representation (Eq.~\ref{eq:mix}), a weighted sum of outer products of complex word embeddings is computed. To ensure the unit trace condition \( \text{Tr}(\rho) = 1 \), weights are normalized into probabilities: \( \lambda_i = \frac{\alpha_i}{\sum_{j=0}^{n}{\alpha_j}} \), where \( \lambda_i \) is the probability of the \( i \)-th word \( \psi_i \). For comparison, we also consider a uniform weighting with \( \lambda_i = 1/n \). In contrast, the superposition-based sentence embedding \( \ket{S} \) directly produces a normalized density matrix (Eq.~\ref{eq:sup}), inherently satisfying the trace condition.

\subsection{Joint Representation}
\label{subsec:joint}
The query and parent entities are represented as density matrices \(\rho_q\) and \(\rho_p\). Instead of using distance-based scoring or concatenation, we model their interaction via a joint representation defined as \(M_{pq} = \rho_q \rho_p\).

We decompose the query and parent density matrices as \(\rho_q = \sum_i \lambda_i \ket{v_i}\bra{v_i}\) and \(\rho_p = \sum_j \lambda_j \ket{v_j}\bra{v_j}\), where \(\lambda_i, \ket{v_i}\) and \(\lambda_j, \ket{v_j}\) are the eigenvalues and eigenvectors of the query and parent respectively. These eigenvectors represent latent concepts (or sememes), weighted by their corresponding eigenvalues. The joint representation is given by \(\rho_q \rho_p = \sum_{i,j} \lambda_i \lambda_j \bra{v_i}{v_j}\rangle \ket{v_i}\bra{v_j}\), where the inner product \(\bra{v_i}{v_j}\rangle\) captures the alignment between basis vectors. Since \(\bra{v_i}{v_j}\rangle = \text{Tr}(\ket{v_i}\bra{v_j})\), the trace inner product becomes \(\text{Tr}(\rho_q \rho_p) = \sum_{i,j} \lambda_i \lambda_j \langle v_i | v_j \rangle^2\), representing a cosine-similarity-based measure between latent spaces. This quantum-inspired similarity, denoted as \(M_{pq}\), encodes the coherence between query and parent \cite{balkir2014using}.

\subsection{Entangled Features for Scoring}

In quantum natural language processing, entity similarity is often measured using the negative von Neumann (VN) divergence, \(-\Delta_{VN}(\rho_p \| \rho_q) = \text{Tr}(\rho_p \log \rho_q)\). However, the matrix logarithm makes it difficult to use in end-to-end learning. To overcome this, we adopt the trace inner product, previously used for word and sentence similarity \cite{blacoe2013quantum}, and shown to approximate the negative VN divergence effectively \cite{sordoni2014learning, zhang2018end}. Formally, it is defined as \(x_{\text{trace}} = \text{Tr}(\rho_q \rho_p) = \sum_{i,j} \lambda_i \lambda_j \langle r_i | r_j \rangle^2\). This expression captures the semantic overlap driving the similarity between the density matrices of the parent and query entities. To enrich their joint representation, we include the diagonal elements of the similarity matrix \( M_{pq} \), denoted as \( \vec{x}_{\text{diag}} \), which reflect varying importance scores. The final feature vector is thus defined as \(\vec{x}_{\text{feat}} = [x_{\text{trace}}; \vec{x}_{\text{diag}}]\). Leveraging these entangled features, we learn a scoring function to effectively rank anchor nodes $n_p \in N^o$ for a query node $q$. We define the scoring function as \(f(\cdot): \mathbb{R}^{D_2} \times \mathbb{R}^{D_1}\rightarrow \mathbb{R}
f^{(i)} = \gamma\left( \mathbf{W}_i f^{(i-1)} + \mathbf{b}_i \right)\) and \(\quad f^{(0)} = \vec{x}_{\text{feat}}
f(n_p, n_q) = f^{(N)} = \sigma\left(f^{(N-1)}\right)\), where $\mathbf{W}_1, \mathbf{W}_2, \mathbf{b}_1$ and $\mathbf{b}_2$ are learnable parameters, $\gamma$ and $\sigma$ are the ReLU and sigmoid activations.

\subsection{Model Training and Inference}
\label{subsec:traininfer}

\textbf{Self-supervised Data Generation.} We use the seed taxonomy $\mathcal{T}^o = (\mathcal{N}^o, \mathcal{E}^o)$ to construct training data in a self-supervised manner. For each edge $\langle n_p, n_c \rangle \in \mathcal{E}^o$, where $n_p$ is the parent and $n_c$ the query term, we create a positive sample $\langle n_p, n_c \rangle$. To generate negatives, we fix $n_c$ and randomly sample $N$ non-descendant nodes $\{n_{p{\prime}}^{l}\}_{l=1}^{N}$ from $\mathcal{N}^o$—typically siblings, cousins, or other relatives of $n_c$. This yields a training instance $\mathbf{X} = \{\langle n_p, n_c \rangle, \langle n_{p{\prime}}^{1}, n_c \rangle, \dots, \langle n_{p{\prime}}^{N}, n_c \rangle\}$. Repeating this for each edge in $\mathcal{T}^o$ forms the self-supervised dataset $\mathbb{X} = \{\mathbf{X}_1, \dots, \mathbf{X}_{|\mathcal{E}^o|}\}$.

\textbf{Model Training.} We train the scoring function $f(\cdot)$ on the dataset $\mathbb{X}$ using the binary cross-entropy loss \(\mathcal{L}(\Theta) = -\sum_{\mathbf{X}_1}^{\mathbf{X}_{|\mathcal{E}^o|}} \sum_{i=1}^{N+1} \Big[y^i \log f\left(\vec{x}^i_{\text{feat}}\right) + \left(1 - y^i\right) \log \left(1 - f\left(\vec{x}^i_{ \text{feat}}\right)\right)\Big]\), where each sample $(\vec{x}^i_{\text{feat}}, y^i)$ corresponds to a candidate pair $\langle n^i_p, n^i_c \rangle$ in data point $\mathbf{X}_k$, with $y^i = 1$ for positives and $y^i = 0$ for negatives.

\textbf{Inference.} Given a query node \(c \in \mathcal{C}\), the goal during inference is to predict its parent node \(n_p \in \mathcal{N}^o\) from the seed taxonomy. For each candidate \(n_p\), we compute a matching score \(f(n_p, n_c)\) and select the parent that maximizes this score: \(n_p := \arg\max_{n_p \in \mathcal{N}^o} f(n_p, n_c)\). Candidates are ranked by their scores, and the top-ranked node is chosen as the predicted parent. This approach can be extended to return the top-$k$ candidates, if needed.

\textbf{Computational Complexity Analysis.} During the training phase, the model processes $|\mathcal{E}^o| \times (N+1)$ training instances per epoch, resulting in a computational cost that scales linearly with the number of edges in the seed taxonomy $\mathcal{T}^o$. During inference, for each query node $ n_c \in \mathcal{N}^o$, the model computes $|\mathcal{N}^o|$ matching scores, one for every node in $\mathcal{N}^o$. Although $O(|\mathcal{N}^o|)$ computation per query can be expensive, it is significantly optimized by computing the scores in batches and accelerating matrix multiplication using GPU resources.

\section{Experimental Setup}
\label{sec:experiments}

We first outline the experimental setup to evaluate the performance of \modelname, covering benchmark datasets, baseline methods, evaluation metrics and implementation details.

\subsection{Benchmark Datasets}
\label{subsec:benchmark}
We evaluate \modelname\ on five benchmarks (c.f. Table~\ref{table:dataset}) -- two SemEval-2016 Task 13 datasets, \textbf{SemEval-Env} (Env) and \textbf{SemEval-Sci} (Sci)~\cite{bordea-etal-2016-semeval}, \textbf{WordNet}, which contains 114 depth-3 sub-taxonomies with 10–50 nodes each~\cite{bansal-etal-2014-structured}, and two large-scale taxonomies, \textbf{MAG-CS-WIKI} (CS) and \textbf{MAG-PSY-WIKI} (PSY), derived from subgraphs of the Microsoft Academic Graph (MAG)~\cite{microsoft}. Concept definitions are taken from prior work~\cite{yu_steam_2020, jiang2023single, wang2021enquire, liu2021temp, taxocomplete}. Following~\cite{yu_steam_2020}, we randomly sample 20\% of leaf nodes as test nodes (with valid parents) and use the remaining nodes as the seed taxonomy for self-supervised training. Since some taxonomies (e.g. CS and PSY) allow multiple parents, but Wu\&P similarity assumes a tree, we derive a spanning tree by duplicating subtrees for multi-parent nodes, ensuring a cycle-free structure with a unique ancestor path for each node.

\subsection{Baseline Methods} 
\label{subsec:baselines}
We evaluate \modelname\ against a range of baselines based on classical embeddings utilizing pretrained language models, graph neural networks, and a few prompting-based methods, all of which encode semantic and structural features. We use the following baselines:
$\bullet$ \textbf{BERT+MLP} \cite{devlin2018bert} encodes term surface forms with \textsc{BERT} and feeds the resulting vectors to a multi-layer perceptron to classify hypernym relationship. $\bullet$ \textbf{TAXI} \cite{panchenko2016taxi} extracts candidate hypernyms via lexical patterns and substring heuristics, then prunes them to produce an acyclic taxonomy without ranking. $\bullet$ \textbf{Arborist} \cite{manzoor2020expanding} models heterogeneous edge semantics and trains with a large-margin ranking loss whose margin adapts dynamically. $\bullet$ \textbf{TaxoExpan} \cite{shen2020taxoexpan} represents an anchor node by encoding its ego network with a graph neural network and scores parent–child pairs through a log-bilinear feed-forward layer. $\bullet$ \textbf{TMN} \cite{zhang2021taxonomy} fuses auxiliary and primary signals using a neural tensor network and refines embeddings via a channel-wise gating mechanism. $\bullet$ \textbf{STEAM} \cite{yu_steam_2020} ensembles graph, contextual, and lexical-syntactic features in a multi-view co-training framework to score hypernymy links. $\bullet$ \textbf{BoxTaxo} \cite{jiang2023single} learns box embeddings and evaluates parent candidates with geometric and probabilistic losses derived from hyper-rectangle volumes. $\bullet$ \textbf{Musubu} \cite{takeoka-etal-2021-low} fine-tunes language-model classifiers on Hearst-pattern phrases that contain both query and parent terms. $\bullet$ \textbf{TEMP} \cite{liu2021temp} encodes full root–to–parent paths and trains with a dynamic-margin loss to represent query nodes.  

Baselines such as TMN are designed for taxonomy completion. They insert a query $q$ between a parent–child pair and therefore score triplets $f(p,c,q)$. Our setting is taxonomy expansion, where q is attached as a leaf under a parent and no child $c$ is available. For a fair comparison, we follow \cite{wang_qen_2022} and adapt these baselines by instantiating $c$ with a dummy placeholder (e.g., a blank/sentinel token). This converts their triplet scorer to an effective leaf-attachment scheme while preserving their original scoring function.

\begin{figure}[!t]
\centering
\includegraphics[width=0.99\columnwidth]{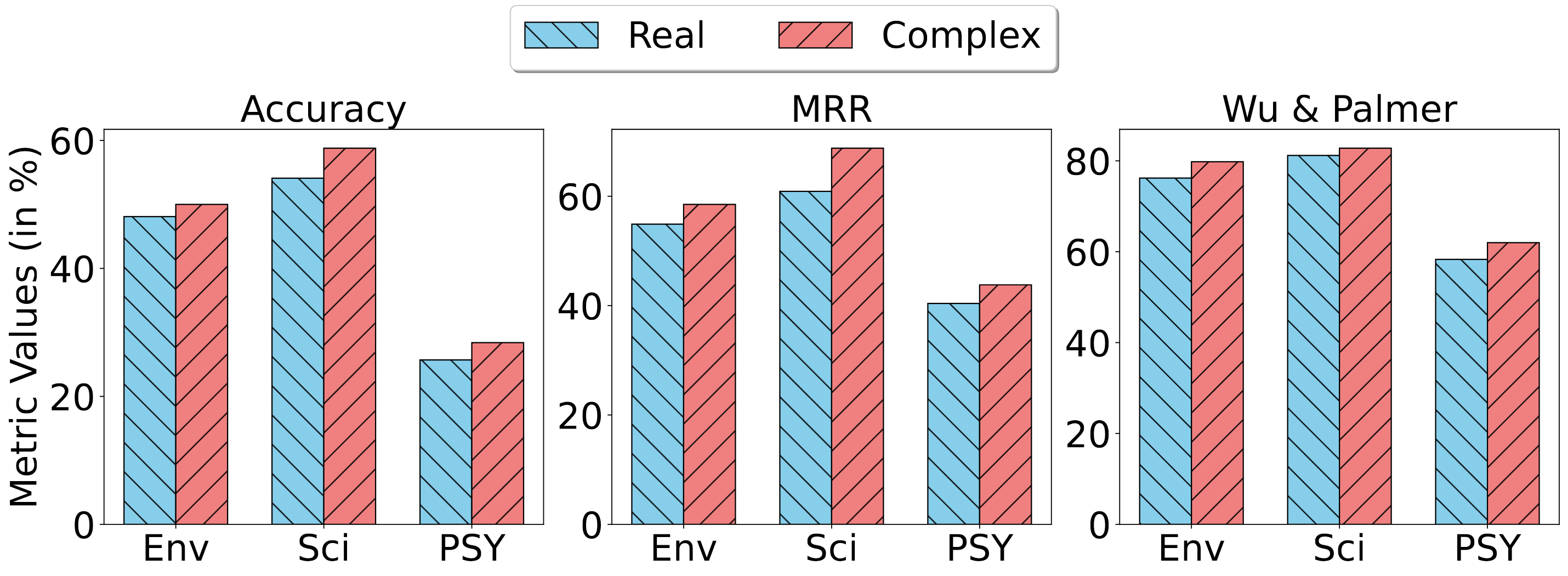}
\caption{Performance comparison of real and complex-valued embeddings across \textit{`Env'}, \textit{`Sci'} and \textit{`PSY'}.}
\label{fig:realvcomplex}
\end{figure}

\begin{figure}[!t]
\centering
\includegraphics[width=1.0\columnwidth]{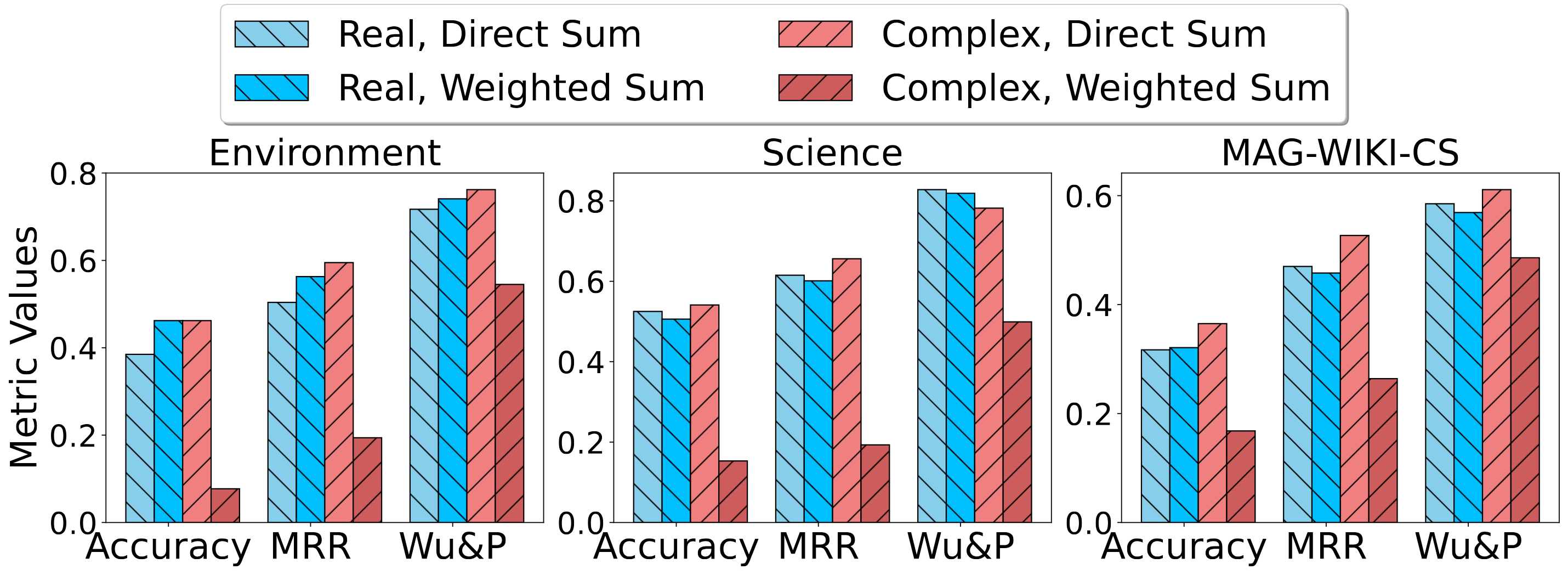}
\caption{Performance comparison of direct and weighted sum in real-valued and complex-valued mixture models across \textit{`Env'}, \textit{`Sci'}, and \textit{`WIKI'} benchmarks.}
\label{fig:directweighted}
\end{figure}

\subsection{Evaluation Metrics}
Given a query set \(\mathcal{C}\), let \(\{\hat{y}_1, \hat{y}_2, \dots, \hat{y}_{|\mathcal{C}|}\}\) be the predicted parent nodes and \(\{y_1, y_2, \dots, y_{|\mathcal{C}|}\}\) the corresponding ground-truth labels. Following prior work \cite{jiang2023single, liu2021temp}, we evaluate the performance of \modelname\ and all baselines using three standard metrics, namely, \textbf{Accuracy (Acc)}\(= \text{Hit@1} = \frac{1}{|\mathcal{C}|} \sum_{i=1}^{|\mathcal{C}|} \mathbb{I}(y_i = \hat{y}_i)\), where \(\mathbb{I}(\cdot)\) is the indicator function, \textbf{Mean Reciprocal Rank (MRR)}\(=\frac{1}{|\mathcal{C}|} \sum_{i=1}^{|\mathcal{C}|} \frac{1}{\text{rank}(y_i)}\) and \textbf{Wu \& Palmer Similarity (Wu\&P)} \cite{wu_verbs_1994}\(= \frac{1}{|\mathcal{C}|} \sum_{i=1}^{|\mathcal{C}|} \frac{2 \times \texttt{DEPTH}(\text{LCA}(\hat{y}_i, y_i))}{\texttt{DEPTH}(\hat{y}_i) + \texttt{DEPTH}(y_i)}\), where \(\texttt{DEPTH}(\cdot)\) denotes the node's depth in the taxonomy. 

\subsection{Implementation Details}
\label{sec:implement}

We implement \modelname\ in PyTorch, with baselines, excluding BERT+MLP, sourced from the respective repositories of their original authors. All training and inference tasks are conducted on an 80GB NVIDIA A100 GPU to ensure high computational efficiency. For the implementation, we utilize \texttt{bert-base-uncased} as the default pre-trained model, with the hidden layer size $W_2$ set to 64 and a dropout rate of $0.1$. The maximum padding length for inputs is fixed at $128$, and the optimizer used is AdamW, with a learning rate of $2 \times 10^{-5}$ for BERT fine-tuning and of $1 \times 10^{-3}$ for training the remaining weights. Training is performed using a batch size of 128 over a maximum of 100 epochs. The sizes of density matrices are kept the same as the dimensions of $\texttt{bert-base-uncased}$, i.e., 768.

\section{Experimental Results}
\label{sec:result}

\subsection{Main Results}
Table~\ref{table:results} shows that \modelname\ consistently outperforms prior state-of-the-art models based on classical embeddings and structural summaries across all evaluation metrics. Early methods like BERT+MLP rely on surface-level names and classical embeddings, ignoring structural cues. First-generation models (e.g., Musubu, TAXI) add lexical and semantic features but struggle with semantic ambiguity. Second-generation methods (e.g., TaxoExpan, STEAM) include structural signals yet still use classical embeddings, limiting semantic depth. Recent baselines like TEMP use path-based transformers but remain bound to classical embeddings. BoxTaxo explores geometric embeddings but is limited by instability and traditional representations. \modelname\ surpasses these baselines by leveraging quantum embeddings to capture semantic entanglements without explicit structural inputs. Its strong performance, even on larger datasets, highlights the representational power and robustness of quantum semantics in modeling taxonomic hierarchy and coherence.

\begin{figure}[!t]
\centering
\includegraphics[width=1.0\columnwidth]{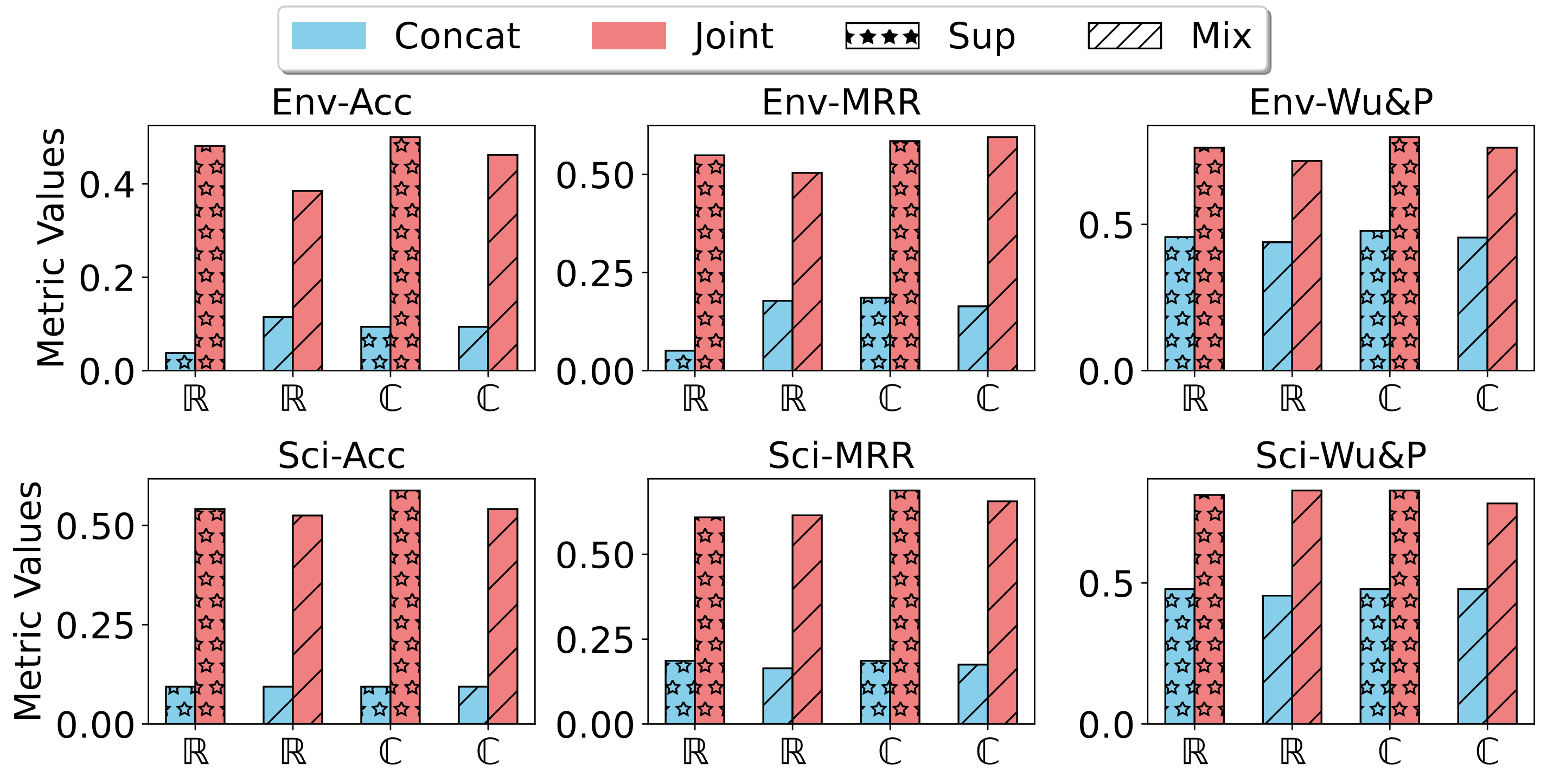}
\caption{Comparison of parent-child joint representation and concatenation for scoring on \textit{`Env'} and \textit{`Sci'}. `Sup' $\rightarrow$ superposition module, while `Mix' $\rightarrow$ mixture module. $\mathbb{R}$ $\rightarrow$ real embedding while $\mathbb{C}$ $\rightarrow$ complex embedding.}
\label{fig:concatvjoint}
\end{figure}

\subsection{Ablation Studies}
\modelname\ comprises three key modules: a complex embedding projector, a quantum module, and a joint representation module. We investigate the effect of different configurations on the performance of taxonomy expansion.

\subsubsection{Impact of Real and Complex Representations on Quantum Modeling.}
We assess the effectiveness of complex-valued embeddings in \modelname\ by comparing them to real-valued embeddings that use only amplitude and omit phase information. As shown in Figure~\ref{fig:realvcomplex}, complex embeddings consistently outperform real-valued ones across all metrics. By incorporating both amplitude and phase, they enable richer, context-aware representations via quantum interference, leading to improved accuracy, higher MRR through better parent ranking, and superior Wu \& Palmer similarity by preserving hierarchical structure. These results highlight the importance of amplitude-phase interactions in capturing nuanced relational semantics, making complex embeddings integral to the success of \modelname.

\subsubsection{Impact of Direct Sum vs. Weighted Sum in Mixture Models.}
We compare direct and weighted summation in real- and complex-valued mixture models across three benchmarks (Section~\ref{subsec:quant}, Figure~\ref{fig:directweighted}). While both methods perform similarly in the real space, weighted summation significantly underperforms in the complex space. This is due to its disruption of the balance between amplitude and phase, which are crucial for capturing semantics in complex embeddings \cite{monning2018evaluation}. Misaligned phases can cause destructive interference, distorting the representation and leading to information loss.

\subsubsection{Comparison of Parent-Child Joint Representation and Concatenation for Scoring.}
We compare joint representation against concatenation for parent-child scoring on the SemEval16-Env and SemEval16-Sci benchmarks (Fig.~\ref{fig:concatvjoint}). Across all settings, joint representation consistently outperforms concatenation. Notably, complex-valued embeddings with joint representation yield the best results, effectively capturing hierarchical and semantic relationships. While real-valued embeddings also benefit from the joint setup, they remain slightly behind their complex counterparts. These findings highlight the superiority of joint representation—especially with complex embeddings—over simple concatenation for modeling relational structure.

\subsubsection{Effect of Dimensionality of Density Matrix on Performance.}
\begin{figure}[!t]
\centering
\includegraphics[width=1.0\columnwidth]{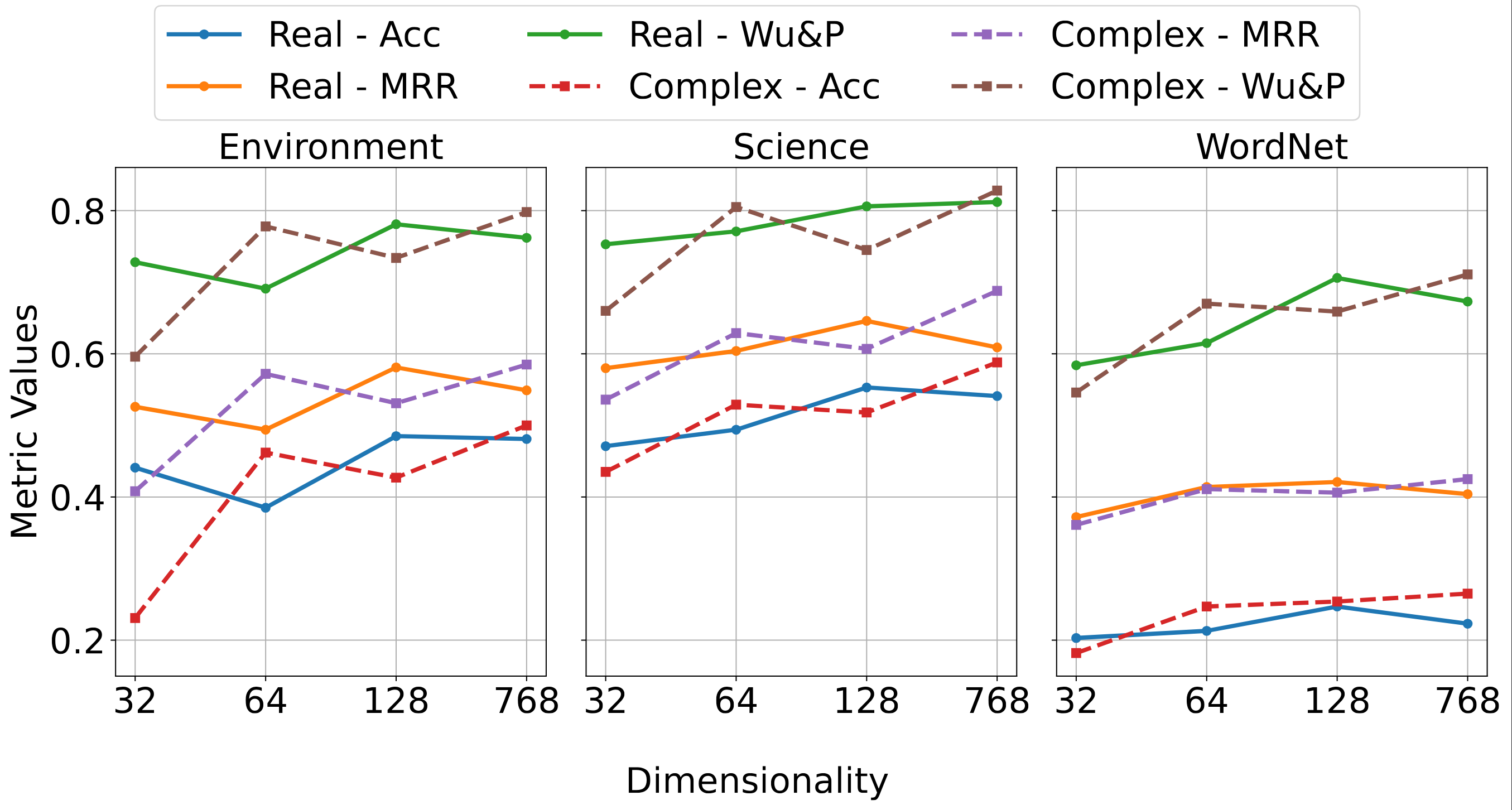}
\caption{Effect of density matrix dimensionality on model performance across  \textit{`Env'}, \textit{`Sci'} and \textit{`CS'}.}
\label{fig:dimensionality}
\end{figure}

We analyze the impact of density matrix dimensionality on model performance (Fig.~\ref{fig:dimensionality}) by projecting BERT embeddings into varying dimensions and computing their outer products. Performance generally improves with higher dimensionality. However, for complex-valued embeddings, performance stabilizes early—for instance, results at dimension 64 closely match those at 768. In contrast, real-valued embeddings continue to benefit from increased dimensions. These findings underscore the efficiency and robustness of complex-valued embeddings in capturing hierarchical and relational semantics, even at lower dimensions.

\begin{figure}[!t]
\centering
\includegraphics[width=0.47\textwidth]{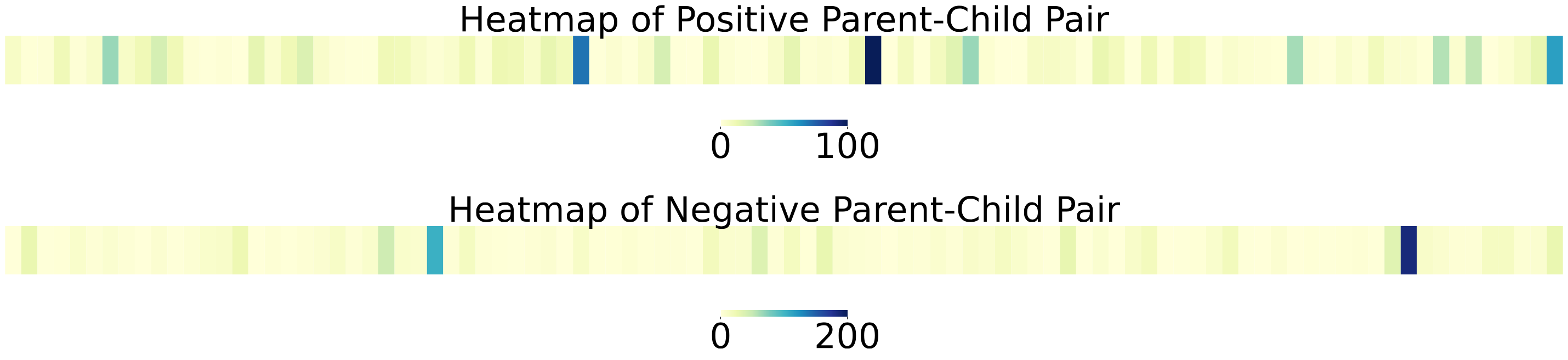}
\caption{Heatmaps of diagonals of joint representation of positive \(\langle\) \textit{"environmental impact"},\textit{"environmental policy"} \(\rangle\) and negative \(\langle\)\textit{"environmental impact"},\textit{"environmental research"} \(\rangle\) pairs, from SemEval16-Env.}
\label{fig:diag}
\end{figure}

\subsection{Distribution of Density Matrices}
We analyze the diagonal features $\vec{x}_{\text{diag}}$ from the joint representation to assess parent-child relationship scoring. Using the SemEval16-Env dataset and the query term \textit{environmental impact} (true parent: \textit{environmental policy}), we compare its density matrix with that of the lowest-scoring negative parent, \textit{environmental research}. As shown in Figure~\ref{fig:diag}, positive pairs exhibit higher intensity and more uniform diagonal values (ranging 0–100), reflecting strong semantic and hierarchical alignment. In contrast, negative pairs show lower intensity, irregular patterns, and sparse high values, indicating weak or absent coherence. These patterns highlight the effectiveness of the joint representation in capturing and distinguishing meaningful parent-child relationships.

\subsection{Case Study}
\label{sec:case}
In this section, we present an error analysis with case studies to evaluate the effectiveness of the \modelname\ framework on SemEval-2016 datasets (Table~\ref{table:case}). For each benchmark, we examine two correct and two incorrect predictions, leading to two key insights on the strengths of quantum embeddings in modeling semantic and hierarchical relationships. \modelname\ performs well on clear and well-defined concepts such as \textit{groundwater}, \textit{tropical zone} and \textit{microbiology}, accurately retrieving appropriate anchor nodes due to the clarity of their definitions. In contrast, it struggles with ambiguous or underspecified terms like \textit{enzymology}, and \textit{non-polluting vehicle}, where insufficient definitions hinder precise semantic grounding. Notably, incorrect predictions generally yield low matching scores, indicating that quantum embeddings avoid forming spurious entanglements in uncertain cases. These lower scores can serve as a heuristic for identifying potentially unreliable predictions in the absence of ground truth.

\renewcommand{\arraystretch}{1.2}

\begin{table}[!t]
\centering
\resizebox{\linewidth}{!}{
\begin{tabular}{|l|l|l|l|c|}
\toprule
\textbf{Dataset} & \textbf{Query} & \textbf{Ground Truth} & \textbf{Predicted Parent} & \textbf{Score} \\
\midrule

\multirow{4}{*}{Env} 
& dust & \makecell[l]{atmospheric\\ pollutant} & environment & 0.809 \\
& \makecell[l]{non-polluting\\ vehicle} & \makecell[l]{pollution control\\ measures} & environment & 0.825 \\
& groundwater & water & water & 0.999 \\
& tropical zone & climatic zone & climatic zone & 0.999 \\

\midrule

\multirow{4}{*}{Sci} 
& radiobiology & biology & science & 0.961 \\
& enzymology & biochemistry & genetics & 0.978 \\
& microbiology & biology & biology & 0.999 \\
& nuclear physics & physics & physics & 1.000 \\

\bottomrule
\end{tabular}
}
\caption{Examples of \modelname's predictions with confidence scores across benchmarks.}
\label{table:case}
\end{table}

\section{Conclusion}
\label{sec:conclusion}

We propose \modelname, a taxonomy expansion framework that uses quantum embeddings to model hierarchical and semantic structure. It consists of three core components: a complex embedding projector, a quantum representation module, and a joint representation module. By projecting entities into a complex-valued Hilbert space, \modelname\ captures rich relationships through superposition and entanglement, which are well-suited for modeling hierarchy and polysemy. The quantum module encodes nuanced semantic overlaps, while the joint module fuses parent and child density matrices for coherent representation. Experiments on benchmarks show that \modelname\ outperforms classical embedding methods and structure-aware models. Ablation studies highlight the impact of quantum representations and complex embeddings in enhancing semantic modeling. Error analysis further reveals \modelname's robustness in handling ambiguous entities. Overall, \modelname\ demonstrates the promise of quantum-inspired approaches for scalable and accurate taxonomy modeling.

\section*{Acknowledgement}
We gratefully acknowledge support from the Prime Minister's Research Fellowship (PMRF), which funded this research.

\bibliography{aaai2026}

\end{document}